\documentclass[preprint,12pt,groupedaddress]{elsarticle}

\usepackage{amssymb}
\usepackage{amsmath}
\usepackage{amsthm}
\usepackage{soul}
\usepackage{graphicx, caption, subcaption}
\usepackage[dvipsnames]{xcolor}

\setul{}{2pt}
\setstcolor{red}

\DeclareGraphicsExtensions{{.eps,.pdf,.png}}
\graphicspath{{./figures/}} 

\journal{Chemical Physics}

\begin{document}

\begin{frontmatter}



\title{Full molecular dynamics simulations of liquid water and carbon tetrachloride for two-dimensional Raman spectroscopy in the frequency domain}

\author{Ju-Yeon Jo}
\ead{ju8879@kuchem.kyoto-u.ac.jp}

\author{Hironobu Ito}
\ead{h.ito@kuchem.kyoto-u.ac.jp}

\author{Yoshitaka Tanimura}
\ead{tanimura@kuchem.kyoto-u.ac.jp}

\address{Department of Chemistry, Graduate School of Science, Kyoto University, Kyoto 606-8502, Japan}

\begin{abstract}
Frequency-domain two-dimensional Raman signals, which are equivalent to coherent two-dimensional Raman scattering (COTRAS) signals, for liquid water and carbon tetrachloride were calculated using an equilibrium-nonequilibrium hybrid MD simulation algorithm. We elucidate mechanisms governing the 2D signal profiles involving anharmonic mode-mode coupling and the nonlinearities of the polarizability for the intermolecular and intramolecular vibrational modes.
The predicted signal profiles and intensities can be utilized to analyze recently developed single-beam 2D spectra, whose signals are generated from a coherently controlled pulse, allowing the single-beam measurement to be carried out more efficiently.
\end{abstract}


\end{frontmatter}


\section{Introduction}

Fifth-order two-dimensional (2D) Raman spectroscopy is the oldest
multi-dimensional laser spectroscopy, whose spectra are obtained by
recording the signals as functions of the time durations between
trains of laser pulses~\cite{Inhomo01}. It has created the possibility for
quantitatively investigating the intermolecular vibrational
motion of liquid molecules, while 2D infrared (IR) spectroscopy was
developed for intramolecular vibrational modes of liquids and
biological molecules~\cite{Hamm,Cho, 2DIR-T03}. Although the possibility of applying 2D Raman
spectroscopy to elucidate the difference between homogeneous and
inhomogeneous broadening~\cite{Inhomo01}, the anharmonicity of the potential~\cite{AH01,AH02,AH03,AH04,AH05}, the
mode-mode coupling mechanism~\cite{couple01,couple02,couple03,couple04,couple05}, and dephasing processes@detected through nonlinear polarizability~\cite{dephase01,dephase02,dephase03,dephase04,dephase05,dephase06,dephase07,dephase08,dephase09,tani2015} of the intermolecular modes were recognized early on, experimental signals have been obtained only recently for CS$_2$~\cite{CS2-01,CS2-02,CS2-03,CS2-04} benzene~\cite{C6H6-01}, and formamide liquids~\cite{HCONH2-01}, due to the unforeseen cascading effect of light emissions~\cite{cascade01,cascade02,cascade03}. While the 2D Raman signals of liquid
water have not yet been observed, 2D THz-Raman (or 2D Raman-THz)
spectroscopy was considered both theoretically and experimentally 
~\cite{2DTR-MD01,2DTR-MD02,2DTR-MD03,2DTR-EX01, 2DTR-EX02,2DTR-MD04,2DTR-MD05,2DTR-TM01}. In this method, the cascading effects are
suppressed using two THz pulses and one set of Raman pulses. Because
intermolecular vibrational modes are usually both Raman and IR
active, the information that we can obtain from the 2D
Raman and 2D THz-Raman signals can be used to investigate the
fundamental nature of intermolecular interactions in a complementary
manner.  Nevertheless, it is desirable to develop 2D Raman
techniques, in which the cascading effects are suppressed, because the
Raman measurement can be applied to a wide variety of materials, including glasses and
solids at high resolution, from low-frequency intermolecular to high frequency intramolecular modes. 

A recently developed single-beam spectrally controlled 2D Raman spectroscopy method, whose signals are
generated from a coherently controlled pulse, overcomes the cascading problem inherent to multi-beam methods and creates a new possibility for measuring intramolecular interactions of liquids by means of 2D Raman spectroscopy~\cite{Frostig}. In this measurement,
a vibrational excitation is created by the sequence of pulses whose time period corresponds to a vibrational mode~\cite{Silberberg1}. Thus, by sweeping the pulse periods, we can tune the excitation frequency of the vibrational modes. 
Although both conventional 2D Raman and single-beam 2D Raman
spectroscopies measure the second-order (or three body) nonlinear
response function of the molecular polarizability, the appearance of the 2D signals is very different. For example,
while the single-beam measurement utilizes a femto-second laser pulse,
the spectrum obtained from this measurement is similar to that obtained from a frequency domain 2D Raman measurement~\cite{couple04}. 
This is because a vibrational excitation in the single-beam measurement is created by a sequence of coherently controlled pulses whose time period corresponds to a vibrational mode, and, by manipulating the two sets of pulse periods, we can obtain the 2D Raman signal as functions of the two vibrational frequencies. In this paper, we calculate the frequency-domain 2D Raman signals of intermolecular and intramolecular modes, which are equivalent to coherent two-dimensional Raman scattering (COTRAS) signals~\cite{couple03}, from full molecular dynamics simulations approach, in order to contribute further development to single-beam 2D Raman spectroscopy.

This paper is organized as follows. 
In Sec. 2, we explain the methodology for simulating 1D and 2D Raman signals with full MD simulations. In Sec.3, we investigate the calculated 1D and 2D Raman signals for liquid water and CCl$_4$ to elucidate the mechanisms governing the 2D signal profiles of the intermolecular and intramolecular vibrational modes. Section~4 is devoted to concluding remarks.

\section{Theory}
\subsection{Symmetrized response functions}
The conventional time-domain 1D Raman measurement utilizes a pair of off-resonant pump pulses, $E_1^2(t)$, followed by the probed pulse, $E_f(t)$. The signal is detected as the function of a delay time between the pump and probe. In the 2D Raman case, the signal is generated by two pairs of off-resonant pump pulses, $E_1^2(t)$ and $ E_2^2(t)$, followed by the final probed pulse, $E_f(t)$, as a function of two delay times between the two pump and probe pulses. The optical observables in 1D and 2D Raman measurements are then defined as the third-order and fifth-order polarizations expressed as~\cite{Inhomo01}
\begin{align}
P^{(3)}(t) &= E_f(t)\int_{0}^{\infty}{dt_1}R^{(3)}(t_1)E_1^2(t-t_1) \label{e:3rd_ttime}
\end{align}
and
\begin{align}
P^{(5)}(t) &= E_f(t)\int_{0}^{\infty}{dt_2}\int_{0}^{\infty}{dt_1}R^{(5)}(t_2,t_1)E_2^2(t-t_2)  E_1^2(t-t_2-t_1) ,  \label{e:5th_ttime} 
\end{align}
where $R^{(3)}(t_1)$ and $R^{(5)}(t_2,t_1)$ correspond to the third- and fifth-order response function defined by
\begin{align}
R^{(3)}(t_1) &= \frac{i}{\hbar}\langle[\Pi(t_1),\Pi(0)]\rangle  \label{e:3rd_R}
\end{align}
and
\begin{align}
R^{(5)}(t_2,t_1) &= \Big(\frac{i}{\hbar}\Big)^2\langle[[\Pi(t_1+t_2),\Pi(t_1)],\Pi(0)]\rangle . \label{e:5th_R} 
\end{align}  
While the signals from conventional time-domain Raman spectroscopy are generated from ultrashort Raman pulses, those from single-beam coherently controlled Raman spectroscopy are generated from trains of short pulses produced using a pulse shaping technique from a single broadband pulse~\cite{Frostig, Silberberg1}. In this measurement, a vibrational excitation is created by a sequence of pulses
whose time period corresponds to a vibrational mode. Thus, by sweeping the pulse periods, we can tune the excitation frequency of the vibrational modes. Both single-beam 1D and 2D Raman measurements utilize a broadband Gaussian electric field with a modulated phase mask produced using a pulse shaping technique described by $E(t)$.~\cite{Weiner1317,Weiner20113669}. 
The signals of single-beam 1D Raman and 2D Raman measurements are then expressed in the Fourier representation as
\begin{align}
P^{(3)}[\omega] 
&= \frac{1}{2\pi}\int_{-\infty}^{\infty}{d\omega_1}S^{(3)}[\omega_1]{E}[\omega-\omega_1]\int_{-\infty}^{\infty}{d\Omega{E}[\Omega]{E}[\Omega-\omega_1]} \label{e:3rd_omega}
\end{align}
and
\begin{align}
P^{(5)}[\omega] 
&= \bigg(\frac{1}{2\pi}\bigg)^2\int_{-\infty}^{\infty}{d\omega_2}\int_{-\infty}^{\infty}{d\omega_1}S^{(5)}[\omega_2,\omega_1]{E}\big[\omega-(\omega_2+\omega_1)\big] \notag \\ &\times \int_{-\infty}^{\infty}{d\Omega{E}[\Omega]{E}[\Omega-\omega_2]}\int_{-\infty}^{\infty}{d\Omega'{E}[\Omega']{E}[\Omega'-\omega_1]} , \label{e:5th_omega}
\end{align}
where $f[\omega]$ is the Fourier representation of any function $f(t)$, and
$\Omega$ and $\Omega-\omega_j$ with $j=1$ and $2$ correspond to the pump and Stokes frequencies. In this measurement, the first- and second-order response functions are also expressed in the Fourier space as $S^{(3)}[\omega]$ and $S^{(5)}[\omega_1,\omega_2]$~\cite{couple03,Mukamel}. It should be noted that the second-order response function in the present case has to be symmetrized in terms of two pulse delay times, because the signal-beam measurement is in principle a frequency domain measurement, where the time ordering of pump and probe pulses do not play a role. Thus, we consider the following response functions
\begin{align}
S^{(3)}[\omega] &\equiv R^{(3)}[\omega]  \label{e:R3_sym} 
\end{align}
and
\begin{align}
S^{(5)}[\omega_1,\omega_2] &\equiv \frac{1}{2!}\bigg\{R^{(5)}[\omega_1+\omega_2,\omega_1]+R^{(5)}[\omega_2+\omega_1,\omega_2]\bigg\}.  \label{e:R5_sym}
\end{align}
The above symmetrized response function is equivalent to the observable of a COTRAS measurement~\cite{couple03}. Although the experimentally observed single-beam 1D and 2D Raman signals were obtained from Eqs. \eqref{e:3rd_omega} and \eqref{e:5th_omega} for a single broadband pulse with a phase mask, here we calculate and analyze $S^{(3)}[\omega]$ and $S^{(5)}[\omega_1,\omega_2]$ themselves to explore the basic features of 2D Raman signals in the frequency domain.

\subsection{Equilibrium-Nonequilibrium hybrid MD approach}\label{sub:1D2DMD}

In 1D vibrational spectroscopic approaches employing full MD simulations, we can evaluate the observables easily by calculating the optical properties from an ensemble of molecular trajectories.
The 1D Raman spectrum, $S[\omega]$, is calculated from the response function in time domain, expressed as
\begin{align} 
 R(t) &= \frac{1}{k_BT} \langle {\Pi}_{\rm eq} (t_1)\dot{\Pi}_{\rm eq} (0) \rangle. \label{MD:1DIR}
\end{align}
Here, ${\Pi }_{\rm eq}(t)$ is the polarizability obtained from the equilibrium MD (EMD) trajectories at time $t$ and $\dot{\Pi}_{\rm eq}(0) \equiv (d{\Pi}_{\rm eq}(t)/dt)_{t=0}$.

To calculate 2D signals, we apply the equilibrium-non-equilibrium hybrid MD algorithm~\cite{Hybrid01, Hybrid02, 2DIR-Raman}.
The second-order response function is evaluated as
\begin{align}
R({t_2},{t_1}) &= 
\frac{1}{k_BT E_2^2 {\Delta}t}\langle \dot{\Pi}_{\rm eq} (-{t_1})
\left( {\Pi}_{+{\Pi}(0)}({t_2})-{\Pi}_{-{\Pi}(0)}({t_2}) \right) \rangle, \label{MD:RTT} 
\end{align}
where ${\Delta}t$ is the time step used in integrating the equations of motion, and $ E_2^2$ is a constant that arises from the non-equilibrium MD (NEMD) perturbation. 
The 2D Raman signal is then calculated as follows. 
We first obtain the time derivative of the polarizability, $\dot{\Pi}_{\rm eq}(-{t_1})$, from the equilibrium trajectories at time $t=-t_1$. Then, we evaluate the polarizability ${\Pi}_{+\Pi (0)}({t_2})$ and ${\Pi}_{-\Pi (0)}({t_2})$ at time $t=t_2$ from the non-equilibrium trajectories, which are generated by a perturbation at time $t = 0$, ${\mp}{\Pi}(0){E}_2^2{\delta}(t)$, resulting from the external electric field of the second pair of pulses ${E}_2^2$ acting on the polarizability $\Pi (0)$. The 2D Raman signal is then calculated from Eq.
\eqref{MD:RTT}.

\subsection{Details of MD simulations} \label{details:MD}
The symmetric response functions, $S^{(3)}[\omega]$ and $S^{(5)}[\omega_1,\omega_2]$, are the key to analyze single-beam 1D and 2D Raman signals. We evaluated these functions from a full molecular dynamics approach. The linear response functions were calculated from the equilibrium trajectory data, while the second-order response functions were calculated with the equilibrium-non-equilibrium hybrid MD simulation algorithm~\cite{Hybrid01, Hybrid02, 2DIR-Raman}.

\subsubsection{Liquid water}

We carried out the MD simulations for liquid water using a polarizable water model for intramolecular and intermolecular vibrational spectroscopies (POLI2VS)~\cite{H-Water}, following the same procedure described in Ref.~\cite{2DIR-Raman}. In the MD simulations, each system consisted of 64 molecules in a cubic box with periodic boundary conditions.
The interaction potentials containing the quadrupole moments were cut off smoothly at a distance equal to half the length of the system using a switching function, and the long-range charge-charge, charge-dipole, and dipole-dipole interactions were calculated with the Ewald summation employing tinfoil boundary conditions.
The equations of motion were integrated using the velocity-Verlet algorithm with ${\Delta}t=$ 0.25 fs.
We first performed isothermal NVT simulations with a Nose-Hoover thermostat.
The conditions of the simulation were set such that the average density and temperature were 0.997 g/cm$^3$ and 300 K, respectively.
Then the production runs of the hybrid MD approach were carried out using the equilibrium molecular dynamics (EMD) simulation in an NVE ensemble for the fixed volume and temperature, followed by the non-equilibrium molecular dynamics (NEMD) simulation with the Raman laser field of 8.0 V/\AA.

\subsubsection{Liquid carbon tetrachloride}
We considered both intermolecular and intramolecular interactions for liquid  $\mathrm{CCl}_4$ using the OPLS force field.
Because the intramolecular motions are described using harmonic functions in the OPLS force field, the anharmonicity of intramolecular modes arises only through intermolecular interactions. The simulations included 32 $\mathrm{CCl}_4$ molecules, with periodic boundary conditions in a cubic simulation box. The box length was chosen to 8.647 \AA, in order to reproduce the experimental density; 1.58 $\mathrm{g/cm}^3$~\cite{Chang:1995}. The equations of motion were integrated using the Velocity-Verlet algorithm with time steps of 2 fs. 
We first performed isothermal NVT simulations with a Nose-Hoover thermostat at 298 K. Then the production runs of the hybrid MD approach were carried out using the equilibrium molecular dynamics (EMD) simulation in an NVE ensemble for the fixed volume and temperature, followed by the non-equilibrium molecular dynamics (NEMD) simulation with the Raman laser field of 0.93 V/\AA. We employed the direct reaction field (DRF) method~\cite{Saito:2003bp} with an Ewald summation to calculate the polarizability of the entire system. The atomic polarizabilities of the carbon and chloride atoms are 1.288599 \AA~and 2.40028 \AA; they were chosen such that the molecular polarizability matches the experimental value 10.002 \AA~\cite{Olney199759}. The damping parameter in DRF method for liquid  $\mathrm{CCl}_4$ was chosen to be 1.95163. The third-order and fifth-order symmetrized response functions were then calculated from Eqs.~\eqref{e:R3_sym} and~\eqref{e:R5_sym}.

\section{Results and discussion}
\subsection{1D Raman spectrum}

\begin{figure}[ht!]
\includegraphics[width=\textwidth]{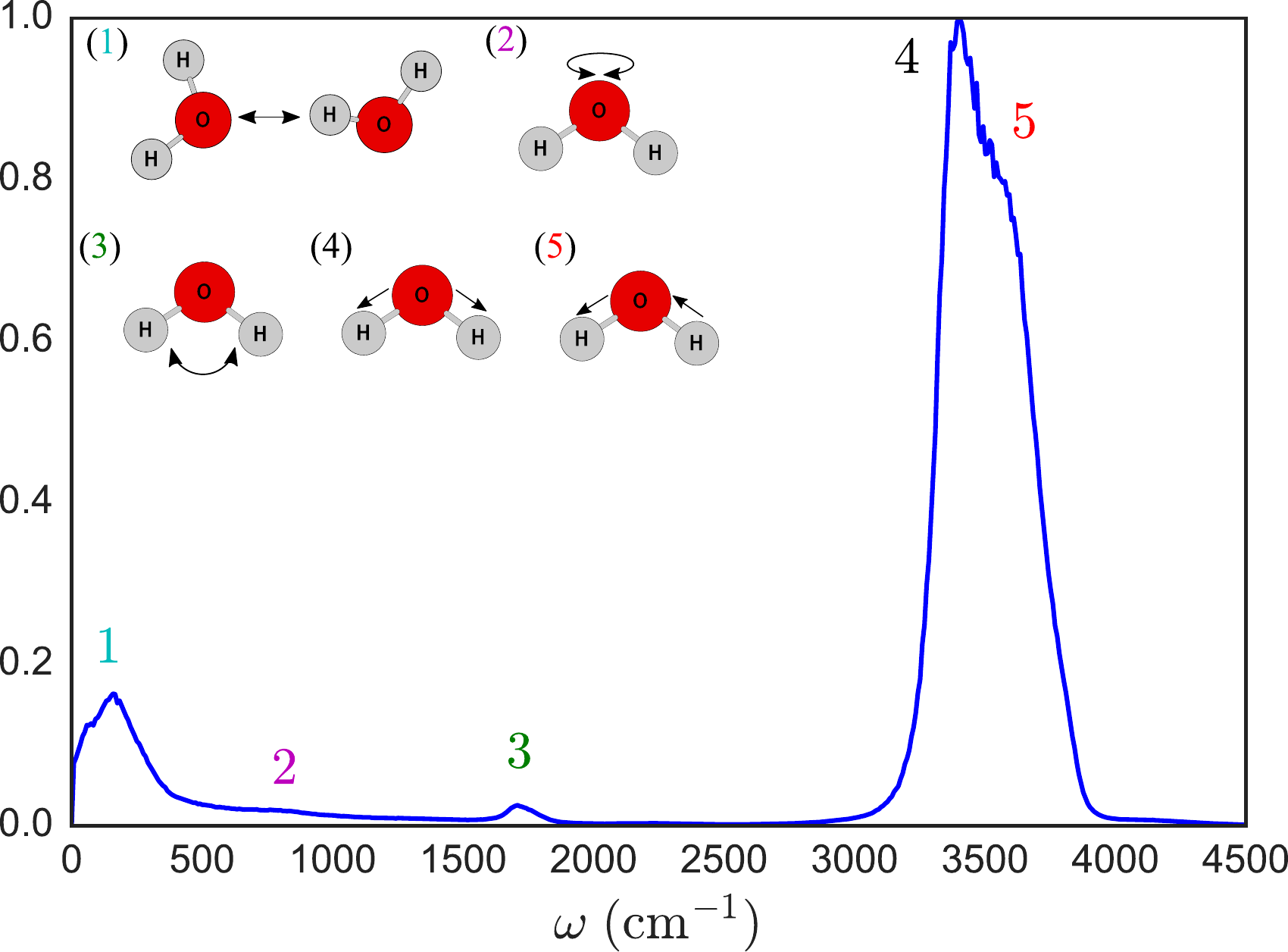} 
\caption{\label{f:1dwatermodes}The 1D Raman spectrum for liquid water obtained from MD simulations. The inset shows the intermolecular and intramolecular vibrational modes of liquid water. The peaks labeled by $n=1$ to 5 arise from these modes.}
\end{figure}

\begin{figure}[ht!]
\includegraphics[width=\textwidth]{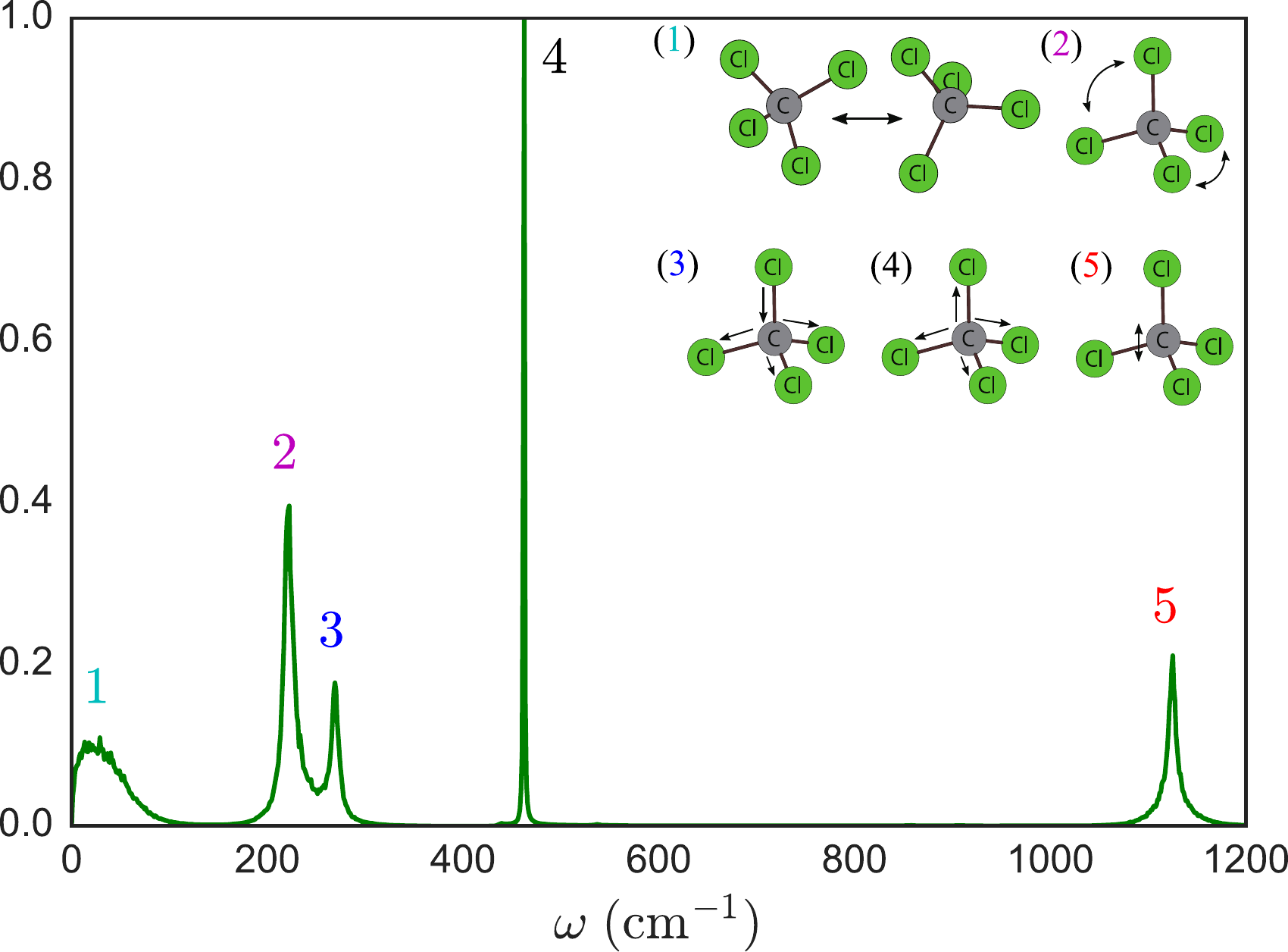} 
\caption{\label{f:1dccl4wmodes}The 1D Raman spectrum for liquid CCl$_4$ obtained from MD simulations. The inset shows the five intermolecular and intramolecular vibrational modes of liquid CCl$_4$. The peaks labeled by $n=1$ to 5 arise from these modes.}
\end{figure}

In Figs.~\ref{f:1dwatermodes} and~\ref{f:1dccl4wmodes}, we show the 1D Raman spectra obtained from MD simulations for liquid water and CCl$_4$. In the case of water in Fig.~\ref{f:1dwatermodes}, the peaks labeled by 1 and 2 arise from the intermolecular translational and librational motion, respectively, whereas the peaks labeled by 3, 4, and 5 arise from the intramolecular bending, symmetric stretching, and anti-symmetric stretching vibrational motion, respectively. Because we calculated the spectrum using the classical MD simulations, the peaks 4 and 5 were blue-shifted \cite{Sakurai-JPCA-2011-115}. In the case of liquid CCl$_4$ in Fig.~\ref{f:1dccl4wmodes}, the peak labeled by 1 arises from the intermolecular interactions between CCl$_4$ molecules, while the peaks labeled by 2 to 5 arise from the intramolecular vibrations. Because the OPLS force field was not accurately designed to reproduce the vibrational frequencies of liquid CCl$_4$, the peaks labeled by 3 and 5 are shifted towards the red and blue by 40 and 345 $\mathrm{cm}^{-1}$, respectively, in comparison with the experimentally obtained spectrum. 

\subsection{Frequency-domain 2D Raman spectrum}
The 2D Raman spectra for liquid water and CCl$_4$ obtained from the MD simulations are presented in Figs.~\ref{f:2dresponse-w} and~\ref{f:2dresponse}.
The top panel in Figs.~\ref{f:2dresponse-w} and~\ref{f:2dresponse} show the double Fourier transform of the 2D Raman signal, $R^{(5)}[\omega_2,\omega_1]$. These spectra are not symmetric, because the nonlinear polarizability involved in the Raman excitations plays a different role depending on the time-ordering of pulses~\cite{dephase03}. 
The bottom panel in Figs.~\ref{f:2dresponse-w} and~\ref{f:2dresponse} show the double Fourier transform of the symmetric 2D Raman signal, $S^{(5)}[\omega_2,\omega_1]$, which corresponds to the COTRAS signal \cite{couple03}. 
It should be noted that the transformation defined by Eq.\eqref{e:R5_sym} does not have a one-to-one correspondence, and we cannot reconstruct $R^{(5)}[\omega_2,\omega_1]$ solely from $S^{(5)}[\omega_1,\omega_2]$. With this we know that the half of the information contained in $S^{(5)}[\omega_1,\omega_2]$ is redundant.  
 Nevertheless, we can estimate the peak positions in $R^{(5)}[\omega_2,\omega_1]$ from experimentally obtained $S^{(5)}[\omega_1,\omega_2]$ using Eq.\eqref{e:R5_sym}.

\subsubsection{Liquid water}

\begin{figure}[ht!]
\centering
\includegraphics[width=0.8\textwidth]{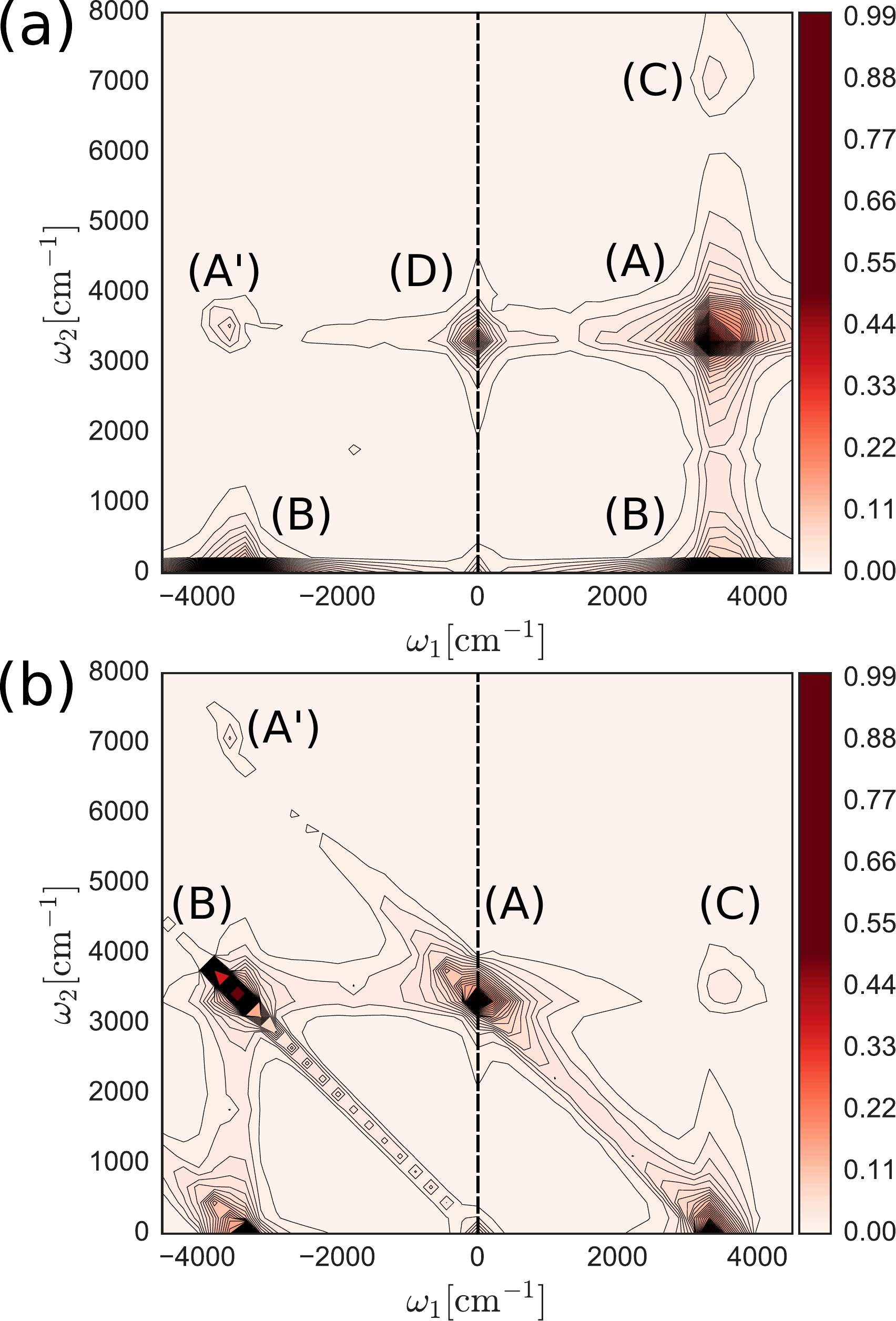}
\caption{\label{f:2dresponse-w}The 2D Raman spectrum for water obtained from MD simulations. (a) The double Fourier transferred spectrum of the second-order response function. (b) The symmetrized spectrum corresponding to the COTRAS measurement.}
\end{figure}
The analysis of the second-order response function, $R^{(5)}[\omega_2,\omega_1]$, for liquid water presented in Fig.~\ref{f:2dresponse-w}(a) has been carried out using a multi-mode Brownian oscillator (BO) model with
nonlinear system-bath interactions \cite{2DTR-TM01,2DIR-Raman} through use of hierarchal Fokker-Planck equations \cite{dephase01,dephase02,dephase03,dephase04,dephase05,dephase06,dephase07,dephase08,dephase09, tani2015}. This result indicates that the peaks labeled by (A)  and (A') arise from the potential anharmonicity, while the peaks labeled by (B) and (C) arise from population relaxation process and the overtone of the OH-stretching mode, respectively. The cross peak labeled by (D) corresponds to the mode-mode coupling between OH stretching mode and intermolecular vibrational mode.
In the symmetrized (COTRAS) case presented in Fig.~\ref{f:2dresponse-w}(b), the peaks labeled by (A), (A'), (B), and (C) in Fig.~\ref{f:2dresponse-w}(a) appear at the location labeled by (A), (A'), (B), and (C) in Fig.~\ref{f:2dresponse-w}(c).
The mode-mode coupling peak labeled by (D) in Fig.~\ref{f:2dresponse-w}(a), however, appears at the same location as the anharmonic peak (A) in  Fig.~\ref{f:2dresponse-w}(b) and may not be separately identified. This is because, unfortunately, the mode-mode coupling peak labeled by (D) is located at $\omega_1 \approx 0$ in Fig.~\ref{f:2dresponse-w}(a), where the symmetric transformation defined by Eq.\eqref{e:R5_sym} does not change location, while the anharmonic peak labeled by (A) in Fig.~\ref{f:2dresponse-w}(a) is transferred to this location. 

\subsubsection{Liquid carbon tetrachloride}

\begin{figure}[ht!]
\centering
\includegraphics[width=0.8\textwidth]{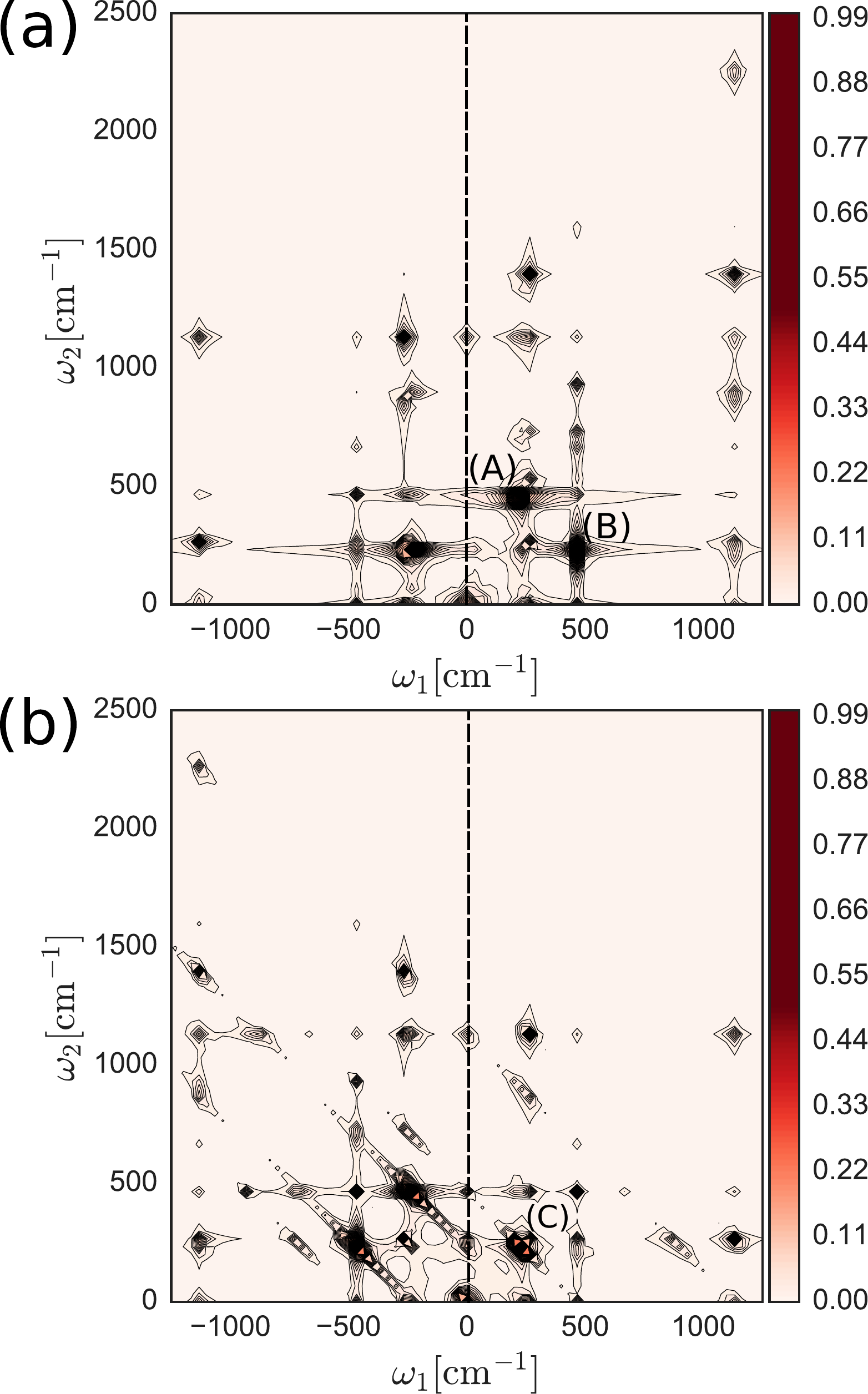}
\caption{\label{f:2dresponse}The 2D Raman spectrum for liquid CCl$_4$ obtained from MD simulations. (a) The double Fourier transferred spectrum of the second-order response function. (b) The symmetrized spectrum corresponding to the COTRAS measurement.}
\end{figure}

In Fig.~\ref{f:2dresponse} (a), we observe the contribution from all five vibrational modes ($\omega = 40, 223, 272, 463, 1125 \mathrm{cm}^{-1}$) as the diagonal, cross, overtone, zero-frequency (axial) peaks, and the sum and difference frequency peaks. As was illustrated in Refs.\cite{2DTR-TM01,2DIR-Raman} for the analysis of molecular modes using a multi-mode Brownian oscillator (BO) model with nonlinear system-bath interactions, these peaks arise from the various types of potential anharmonicity and nonlinearity of polarizability.
Because the OPLS force field for liquid CCl$_4$ assumes harmonic potentials for the intramolecular modes, the main contributions to the signal arise from the nonlinear polarizability of molecular modes. Indeed, the overtone peak of the four intramolecular modes ($\omega = 223, 272, 463, 1125 \mathrm{cm}^{-1}$) as well as their sum and difference frequency peaks are originated from the nonlinear polarizability. It should be noted that the peaks labeled by (A) and (B) in Fig.~\ref{f:2dresponse} (a) are large and broadened, because these peaks consist of peaks from a different origin. For example, some contribution of the peaks labeled by (A) and (B) arise from a pair of two modes, $223$ and $463\mathrm{cm}^{-1}$, and $272$ and $463\mathrm{cm}^{-1}$, respectively. Moreover the peak labeled by (A) contains the overtone peak of $223 \mathrm{cm}^{-1}$ and the sum frequency peaks of $223$ and $272\mathrm{cm}^{-1}$, while the peak labeled by (B) contains the different frequency peaks between $223$ and $463\mathrm{cm}^{-1}$, and between $272$ and $463\mathrm{cm}^{-1}$. 

In the case of the symmetrized (COTRAS) spectrum in Fig.~\ref{f:2dresponse} (b), these two peaks labeled by (A) and (B) in Fig.~\ref{f:2dresponse} (a) transferred through Eq.\eqref{e:R5_sym} to a similar location labeled by (C). Thus, although the peak labeled by (C) is prominent compared with the other surrounding peaks, we cannot solely identify the origin of this peak. To investigate further, we must analyze not only the peak labeled by (C) itself but also the surrounding peaks in a consistent manner. These surrounding peaks may originate from the same vibrational mode, but arise from a different mechanism of anharmonic or nonlinear polarizability.

\section{Conclusion}
In this work, we have calculated frequency-domain 2D Raman signals, which are equivalent to coherent 2D Raman scattering (COTRAS) signals, for liquid water and carbon tetrachloride. We found that the analysis of the frequency-domain 2D Raman spectrum was harder than the analysis of conventional 2D Raman spectrum, because some of information carried by the conventional 2D Raman response functions were lost due to the symmetrization of the response function in frequency domain. In order to analyze frequency-domain 2D signals, it is therefore important to construct a consistent model that explains not only the location and intensity of the targeting peak itself but also those of the surrounding peaks. 
With the help of full molecular dynamics simulation and model based analysis, we expect to be able to elucidate the key features of frequency-domain 2D Raman signals obtained from experiment, with a view towards probing the fundamental nature of intermolecular and intramolecular interactions. 

To analyze spectra obtained from a single-beam 2D Raman measurement, it is important to develop a methodology to simulate and to analyze frequency-domain 2D Raman signals, because the observable of this new measurement is formulated on the basis of the symmetrized 2D Raman response function.  
In order to make a direct comparison between the present results of our simulations and experimentally obtained results, we must calculate the polarizability defined by Eqs.\eqref{e:3rd_omega} and \eqref{e:5th_omega} for the specific form of electric field used for the single-beam measurement. This research direction is left for future studies.

\section*{Acknowledgements}
J. J. acknowledges the stimulating discussion with H. Frostig. 
This research is supported by a Grant-in-Aid for Scientific Research (A26248005) from the Japan Society for the Promotion of Science.

\end{document}